%File: bjb.tex -- macros for BJB         
%
% Page sizes
%
\def\BJBa4{
 \hsize=6.3in
 \vsize=9.6in
 \voffset=-0.3in
}

\def\laeq{\lower.5ex\hbox{\fiverm{\ $\buildrel < \over \sim$\ }}}
\def\gaeq{\lower.5ex\hbox{\fiverm{\ $\buildrel > \over \sim$\ }}}
\def\boxit#1{\vbox{\hrule\hbox{\vrule\kern3pt
     \vbox{\kern3pt#1\kern3pt}\kern3pt\vrule}\hrule}}
%
% Line spacing
%

%
% Horizontal ruling for tables
%

%
% MN pedantry
%

\def\leftdisplay#1$${\leftline{\noindent$\displaystyle{#1}$}$$}
%
%Comment next line for centred equations
%
%\everydisplay{\leftdisplay}
%
%  This paragraph defines the fonts for tw.
%       
 \font\medbx=cmbx10 scaled \magstep2

 \font\twelvebf=cmbx12 
 \font\twelvett=cmtt12 at 12pt
 \font\twelverm=cmr12 at 12pt    \font\ninerm=cmr9
\font\sevenrm=cmr7    \font\fiverm=cmr5
 \font\twelvei=cmmi12 at 12pt    \font\ninei=cmmi9
\font\seveni=cmmi7    \font\fivei=cmmi5
 \font\twelvesy=cmsy10 at 12pt   \font\ninesy=cmsy9
\font\sevensy=cmsy7   \font\fivesy=cmsy5
 \font\twelveit=cmti12 at 12pt
 \font\twelvesl=cmsl12 at 12pt
 \font\tenex=cmex10
 \font\teni=cmmi10    \font\tensy=cmsy10
 \font\tentt=cmtt10   \font\tenit=cmti10
 \font\tensl=cmsl10
%
% twelvepoint fonts
% 
\def\BJBtwelvepoint{\def\rm{\fam0\twelverm}
 \textfont0=\twelverm \scriptfont0=\ninerm \scriptscriptfont0=\sevenrm
 \textfont1=\twelvei \scriptfont1=\ninei \scriptscriptfont1=\seveni
 \textfont2=\twelvesy \scriptfont2=\ninesy \scriptscriptfont2=\sevensy
 \textfont3=\tenex \scriptfont3=\tenex \scriptscriptfont3=\tenex
 \def\it{\fam\itfam\twelveit}%
 \textfont\itfam=\twelveit
 \def\sl{\fam\slfam\twelvesl}%
 \textfont\slfam=\twelvesl 
 \def\bf{\fam\bffam\twelvebf}%
 \textfont\bffam=\twelvebf 
 \def\tt{\fam\ttfam\twelvett} % \tt is family 7
 \textfont\ttfam=\twelvett
 \font\bigit=cmti12 scaled \magstep2
 \baselineskip 14pt%
 \abovedisplayskip 14pt plus 3pt minus 10pt%
 \belowdisplayskip 14pt plus 3pt minus 10pt%
 \abovedisplayshortskip 0pt plus 3pt%
 \belowdisplayshortskip 8pt plus 3pt minus 5pt%
 \parskip 3pt plus 1.5pt
 \setbox\strutbox=\hbox{\vrule height10pt depth4pt width0pt}%
 \rm}
%
% tenpoint fonts
%
\def\BJBtenpoint{\def\rm{\fam0\tenrm}
 \textfont0=\tenrm \scriptfont0=\sevenrm \scriptscriptfont0=\fiverm
 \textfont1=\teni \scriptfont1=\seveni \scriptscriptfont1=\fivei
 \textfont2=\tensy \scriptfont2=\sevensy \scriptscriptfont2=\fivesy
 \textfont3=\tenex \scriptfont3=\tenex \scriptscriptfont3=\tenex
 \def\sl{\fam\itfam\tenit}%
 \textfont\itfam=\tenit
 \def\sl{\fam\slfam\tensl}%
 \textfont\slfam=\tensl 
 \def\bf{\fam\bffam\tenbf}%
 \textfont\bffam=\tenbf 
 \def\tt{\fam\ttfam\tentt}%
 \textfont\ttfam=\tentt
 \baselineskip 12pt%
 \abovedisplayskip 14pt plus 3pt minus 10pt%
 \belowdisplayskip 14pt plus 3pt minus 10pt%
 \abovedisplayshortskip 0pt plus 3pt%
 \belowdisplayshortskip 8pt plus 3pt minus 5pt%
 \parskip 3pt plus 1.5pt
 \setbox\strutbox=\hbox{\vrule height8.5pt depth3.5pt width0pt}%
 \rm}

\BJBa4
\BJBtenpoint
\noindent
%text
\ 
\vglue 2.0cm
\centerline{\medbx Morphology and Surface Brightness Evolution} 
\centerline{\medbx of z$\sim$ 1.1 Radio Galaxies} 
\vskip 1.5cm
\centerline{Nathan Roche, Stephen Eales}
\vskip 0.5cm 
\centerline{Department of Physics and Astronomy,
University of Cardiff,}
\centerline{P.O. Box 918, Cardiff CF2 3YB, Wales.}
\vskip 1.0cm
\centerline{Steve Rawlings}
\vskip 0.5cm 
\centerline{Department of Astrophysics,
University of Oxford,}
\centerline{Nuclear and Astrophysics Laboratory, Keble Road, 
Oxford OX1 3RH, England.}
\vskip 2.5cm
\centerline{\bf Abstract}
\smallskip
We use $K^{\prime}$-band (2.1 $\mu \rm m$) imaging to investigate the angular size and morphology of 10 6C radio galaxies, at redshifts $1\leq z\leq 1.4$. Two appear to be undergoing mergers, 
another contains two intensity peaks aligned with the radio jets, while the
  other seven appear consistent with being normal ellipticals in the $K$-band. 
 
Intrinsic half-light radii are estimated from the areas of each radio galaxy 
image above a series of thresholds. The
6C galaxy radii are found to be significantly smaller than those of the more radioluminous 3CR galaxies at similar redshifts. This would indicate
 that the higher mean $K$-band  luminosity of the 3CR galaxies results from a difference in the size of the host galaxies, and not solely from a difference in the power of the active nuclei. 

The size-luminosity relation of the $z\sim 1.1$ 6C galaxies indicates a 1.0--1.8 mag enhancement of the rest-frame $R$-band surface brightness relative to 
either local ellipticals of the same size or FRII radio galaxies at $z<0.2$.  
The 3CR galaxies at $z\sim 1.1$ show a comparable enhancement
 in surface brightness.  The mean radius of the 6C galaxies suggests that they evolve into ellipticals of $L\sim L^*$ luminosity, and is consistent with their low redshift counterparts being relatively small FRII galaxies, a factor $\sim 25$ lower in radio luminosity, or small FRI galaxies
a factor of $\sim 1000$ lower in radio luminosity. Hence the 6C radio galaxies may undergo 
at least as much optical and radio evolution as the 3CR galaxies.
\bigskip
\noindent {\bf 1. Introduction}
\smallskip
Powerful radio galaxies at $z<0.5$ appear similar in morphology and profile to normal giant ellipticals (Lilly, MacLean and Longair 1983), with luminosities in the range from $L^*$ to
$\sim 2$ mag above $L^*$ and no obvious correlation between radio and optical luminosity (Laing, Riley and Longair 1983; Owen and Laing 1989; 
Owen and White 1991). The relative uniformity of the 
optical properties of radio galaxies, combined with the ease of detection at higher redshifts, means that they provide a useful probe of galaxy evolution.  
Lilly and Longair (1984) measured $K$-band ($2.2\mu \rm m$)
magnitudes for 3CR radio galaxies over  a very wide range of redshifts
(the 3CR catalog is a flux limited sample containing the radio galaxies with apparent flux
exceeding 10 Jy at 178 MHz). The
 $K$--$z$ relation of these galaxies, which showed 
little scatter out to $z\sim 2$,
indicated an evolutionary brightening with redshift,
 consistent with a high formation redshift ($z\sim 5$) followed by passive
stellar evolution.

The radio luminosities of the
most powerful radio galaxies also evolve.
Laing, Riley and Longair (1983), using a $V/V_{max}$ test, showed that their decrease in luminosity with time was
comparable to the evolution of quasars.
Padovani and Urry (1992) studied the radio
evolution of the 3CR catalog in more 
detail, deriving a radio luminosity function and obtaining a best fit for an exponential $L_{rad}$ evolution
with timescale
$\tau=0.17^{+0.03}_{-0.02}H_0^{-1}$ i.e. $\tau=3.3^{+0.6}_{-0.4}$ Gyr.
It must be noted that this exponential decrease describes the evolution of the radio luminosity function, rather than that of individual radio galaxies.
The galaxies might, for example,
undergo short bursts of radio emission, 
separated by much longer periods of radio quiescence, with subsequent
bursts decreasing with time
in their peak radio intensity.

Higher redshift radio galaxies differ in appearence from normal ellipticals,
especially at shorter wavelengths. Rigler et al. (1992) studied  
13  3CR galaxies at $0.8<z<1.3$ and
found that they could be decomposed into
`active' and `passive' components.
The `active' components were elongated, blue
(approximately $f_{\nu}\propto \nu^0$), and 
aligned with the radio axis. Best et al. (1996), using WFPC2 imaging, found that in radio galaxies with a relatively small separation between the radio lobes, the blue components consisted of  strings of several bright knots, whereas in radio galaxies with radio lobes separated by more than $\sim 200$ kpc, the blue components were much
more compact. This suggested an evolutionary sequence in which the elongated optical components were formed as the outward passage of the radio jets triggers large-scale star-formation, and later fall back towards the galaxy centres.  However, the aligned components of at least some 
$z\sim 1$ radio galaxies show some polarization (Tadhunter et al. 1992; 
Leyshon and Eales 1997), suggesting that dust-reflected light from the active nucleus is also
important. 

The aligned components are less prominent at longer wavelengths, typically contributing only $\sim 10\%$ of the total 
flux at $\lambda_{rest}\simeq 1 \mu \rm m$ (i.e. the observed $K$-band), but
 many of the $z\sim 1$ 3CR
galaxies still show some alignment between their radio axes and $K$-band isophotes (Dunlop and Peacock 1993; Ridgway and Stockton 1997).
 Most of the red light is produced by the `passive' components, which are redder, diffuse and symmetric, apparently 
unaligned with the radio axis, and appear to be underlying giant ellipticals.
 
 In the infra-red $H$-band ($1.65\mu \rm m$), Rigler and Lilly (1994) found the profile of the $z=1.18$ radio galaxy 3C65 
to be well-fitted with  a pure bulge ($r^{1 \over 4}$)
profile of effective radius $r_e=1.77\pm 0.18$ arcsec.
Best et al. (1997), observing in the $K$-band and with the HST WFPC2,
 confirmed $r_e=1.7$ arcsec for 3C65, and found all 21 of a sample of $z\sim 1$ 3CR galaxies to be well-fitted by bulge models with 
$0.7<r_e<3.9$ arcsec. The size-luminosity relation of these galaxies was 
consistent with that of of local ellipticals with the 
$\sim 1^m$ of brightening expected from passive evolution.
Only two of the 21 galaxies contained substantial 
($>10\%$ of total flux) nuclear point-sources components, although several showed some evidence of an excess of 
surface brightness above a bulge-model profile
at large radii ($r>35$ kpc), like that in cD galaxies, which might
be evidence for a rich cluster environment. 

In this paper we investigate the $K$-band morphology and radii of 
10 galaxies at similar redshifts but 
with more moderate radio
luminosities, selected from the 6C survey which has a flux limit about six
times fainter than 3CR. There are some unsolved problems concerning the differences between the 3CR and 6C galaxies.  
Firstly, 6C radio galaxies at $0.6<z<1.8$ 
were found to be on average 0.7 
mag less luminous in the observed $K$-band than 3CR galaxies at the same redshifts
(Eales and Rawlings 1996; Eales et al. 1997), but it was unclear
whether this was due solely to a greater near-IR flux from the 3CR nuclei, or to
 a greater  mass of stars in the 3CR galaxies. We shall investigate whether
the differences in the
 radio and optical luminosities of 3CR and 6C galaxies at $z\sim 1$
are related to any
differences in the 
old stellar component visible in the $K$-band. 
Secondly, as the 3CR and 6C galaxies  at lower redshifts are similar
in mean $K$-band luminosity, the $K$--$z$ relations of the 3CR and 6C 
catalogs are different in slope, with the latter consistent with no evolution.
Hence if high redshift 6C galaxies evolve into the
low redshift 6C galaxies, they
undergo surprisingly little change in  $K$-band
luminosity, but it is also possible that   their low redshift counterparts are
galaxies of 
lower $K$ and radio luminosity.
By determining a size-luminosity relation for the 6C galaxies and comparing with both $z\sim 1$ 3CR galaxies and  low-redshift radio galaxies (from Owen and Laing 1989), we may compare the surface brightness evolution of the 
two catalogs and determine whether the 6C galaxies are truly non-evolving.
 
 We assume $H_0=50$
km $\rm s^{-1}Mpc^{-1}$ and $q_0=0.05$ throughout.
Section 2 describes the observational data, Section 3 its analysis 
and photometry. In Section 4 we describe the estimation of the size and morphological type of the galaxies, and in Section 5 discuss the appearence of each individual 6C 
galaxy. In Section 6
we present the size-luminosity relation and compare with 3CR 
galaxies at similar redshifts and local radio galaxies, and in Section 7 discuss the implications for radio galaxy
evolution. Section 8 summarizes the main conclusions.
\bigskip
\noindent {\bf 2. Observational Data}
\smallskip
 The observational dataset of this paper consists of 10 
images, each centred on a 6C radio galaxy with
a known spectroscopic redshift in the $1.0\leq z\leq 1.4$ range
($z_{mean}=1.11$). This data was previously used by
Eales et al. (1997) in a study of  
the $K$-band luminosity evolution of radio galaxies.
 The 6C galaxies considered here have 151 MHz fluxes
of 2.2--4.4 Jy, corresponding at these redshifts to
$L(151 $MHz$)\sim 10^{28.6}$ W $\rm Hz^{-1}$. Allington-Smith (1982) and 
Eales (1985) give
more information on the radio source properties, 
Eales et al. (1997) and references therein detail
the identification of the optical counterparts, and
Rawlings et al. (1997) describe the spectroscopic observations.
For comparison we include in our dataset an image of a
 radio-loud QSO, 5C6.8, which lies in the 
same redshift range as the other objects. Table 1 lists the radio galaxy co-ordinates and redshifts for all 11 images.

These fields were observed over the period 18--20 January 1995 using the Redeye camera on the Canada-France Hawaii Telescope (CFHT). Redeye is
a $256\times 256$ HgCaTe mosaic with a pixelsize of 0.50 arcsec, covering
approximately $2\times 2$ arcmin. Each field was imaged nine times, with the camera being
offset by 8 arcsec between exposures in a $3\times 3$ grid.  The total 
exposure times varied from 26 to 47 minutes per field. 
The observations were
made in the the $K^{\prime}$ band, centred on a wavelength of $2.1\mu \rm  m$ where there  is some reduction of sky background compared to that in the
 standard 2.2 $\mu\rm m$ $K$-band (Wainscoat and Cowie 1992).
All 11 images had good seeing, estimated as $FWHM\simeq 0.9$--1.2
arcsec. 
 \bigskip
\noindent {\bf 3. Data Analysis and Galaxy Detection}
\smallskip
After initial image processing using the IRAF package (see Eales et al. 1997), the  Starlink PISA (Position, Intensity and Shape Analysis) package, developed by M. Irwin, P. Draper and N. Eaton, was used to detect and catalog the
objects on each field. 
Object were detected on the basis that they exceed an intensity threshold of
$1.5\sigma$ above the background noise ($\sigma$ being separately determined by
PISA for each of the frames) in at least 6 connected
pixels. Fluxes were estimated 
using the `total magnitudes' option within PISA, which counts 
photons above the sky background level in elliptical apertures about the
centroid of each detection, with the aperture size and shape being fitted by a
curve-of-growth analysis to the intensity profile of each object.

 Observations of $K$-band standard stars provided a photometric zero-point for
the Redeye data of $K=21.28$ for 1 count $\rm sec^{-1}$. However, due to the 
offset between the $K$ and $K^{\prime}$ passbands,
 this zero point will be exact only for objects with the same
$K^{\prime}-K$ as the standard stars. Radio galaxies with elliptical-like
colours will at $1\leq z\leq 1.4$ be redder than the 
calibration stars by an estimated
$\Delta(K^{\prime}-K)\simeq 0.13$ mag (see Eales et al. 1997), so to correct for this we add 
$\Delta(K)=-0.13$ mag to the magnitudes,
adopting a zero-point of $K=21.15$  for 1 count $\rm sec^{-1}$.

PISA also produces a list of 8 
areas for each detected object. The first area is simply the number of pixels 
where the intensity equals or exceeds
the chosen detection threshold, while sizes 2--8 are the areas above higher
thresholds of $I_{thr}+2^{j+2}$ counts, where $I_{thr}$ is the threshold and
$j$ an integer from 2 to 8. For 
 these images, $I_{thr}\sim 50$ counts so
the 8th threshold would be $\sim 1074$ counts and $\sim 7.5$ mag 
above the detection threshold.

 We are only concerned here with the radio galaxies on each image;
 the properties (e.g. clustering) of the other galaxies detected will be discussed in  a separate paper.
 Figures 1 and 2 show greyscale images and contour plots
of $15\times 15$ arcsec areas centred on each radio galaxy.
Table 1 lists the $K$-band total magnitudes of the radio galaxies, as measured from this 
data (with errors estimated from the best-fitting models described in Section 
4 below). These are consistent with the 
magnitudes of Eales et al. (1997). Table 1 also lists
 the detection thresholds of the 11 images in terms of $K$ mag $\rm arcsec^{-2}$.
Two of the radio galaxies were detected by PISA as
 double objects on these
images, 6C011+36 with components of apparent separations  1.7 arcsec
 and 6C1256+36 with components of separation
  2.5 arcsec. These pairs are still connected at the detection threshold so
are only split when PISA is run with the `deblend' option. In this analysis we have used 
PISA without deblending, so these radiogalaxies are classed as single objects, and the derived 
magnitudes and areas will be those of both components together.

\bigskip
\noindent{\bf 4. Estimating Galaxy Sizes and Profiles}
\smallskip
\noindent{\bf 4.1 Method}
\smallskip
Most of the radio galaxies appear extended on these images.
To  estimate their intrinsic angular sizes and  distinguish
between disk and bulge profiles, we compare their areas above 8 thresholds as given by PISA  with those measured for   
simulated  galaxy profiles.   
 We consider three types
of source profile

(i) A point source.

(ii) An exponential disk, with surface brightness $\mu\propto{\rm exp}(-r/r_{exp}$). We consider
 exponential scale-lengths in the range $0.07\leq r_{exp}\leq 1.00$ arcsec.
The half-light radius $r_{hl}=1.68r_{exp}$.

(iii) A bulge (elliptical galaxy) profile, with surface brightness
 $\mu\propto{\rm exp}-7.67(r/r_{e})^{0.25}$. We consider
effective radii ($r_e$) in the range 
$0.07\leq r_{e}\leq 3.00$ arcsec. For this profile $r_{hl}=r_{e}$.

Each model profile was generated on a 0.1 arcsec pixel grid, and then rebinned into
0.5 arcsec pixels. The model profiles were arranged in 
a grid pattern on a `simulation image', with each model profile
represented 12 times with slightly different positional offsets relative to the
0.5 arcsec pixel grid.
 The seeing profile was estimated for each image 
by fitting (using a PISA routine)  a combined Gaussian-exponential-Lorentzian
to a few objects identified as stars (of similar apparent magnitude to the radio galaxies). The simulation
image  was then convolved with the model stellar profile and normalized so that the model profiles had 
the same total intensity as measured for the radio galaxy. Gaussian noise was then added to the simulation image,
 with $\sigma$ equal to the sky noise as measured on the real data.

The simulation image was then analysed using PISA, with the same detection
threshold as used for the radio galaxy data. For each model profile, 
this gave 12 sets of areas 
above the 8 thresholds, with some scatter between the 12 due to the modelled
 sky noise and the positional offsets between
 the models. These 12 sets of areas were averaged to give a set of 8 
PISA areas corresponding to each model profile, with error bars
 calculated from the  scatter. For each radio galaxy we
 then compared the observed set of 8 
areas with those measured for each of the models,  using a 
$\chi^2$ test with the simulation errors.

Table 2 shows the results of this comparison for each object, with $\chi^2$ for the point-source models,
the exponential
scalelength ($r_{exp}$) and $\chi^2$ of the disk-profile model which best fits the
observed set of 8 areas, i.e. which gives the smallest $\chi^2$, and the effective radius ($r_{e}$) and $\chi^2$ of the best-fitting 
bulge model. 

For a $\chi^2$ test with 7 degrees of freedom, we would expect
a `perfect' model to on average give $\chi^2\simeq 6.4$, and would 
estimate $\pm 1\sigma$
errors on  $r_{exp}$ and $r_{e}$ as the change in the model radius above or below the best-fitting model which increases $\chi^2$ by 8.16 above its minimum value. However, in this analysis the 8 areas given by PISA are not entirely independent; a large noise fluctuation could change a pixel's value by
more than one threshold interval, thus altering more than one of the areas.
To take this into account, we also $\chi^2$-test the observed areas above thresholds against each of the 12 representations of the best-fitting model in our simulation image, obtaining 12 values of $\chi^2$. The scatter
in these 12 values of $\chi^2$ gives the $\pm 1\sigma$ errors (listed in Table 2) produced by noise on the best-fit model's
$\chi^2$. The errors
on the best-fit $r_{e}$ and $r_{exp}$ can be estimated as the change in these radii  which increases $\chi^2$ above that of the best-fit model by these $1\sigma$ noise errors. However, the 
the true errors on the size estimates may be somewhat larger
due to the approximate modelling of the point-spread function and to asymmetry
and substructure in the real radio galaxies.  
\smallskip
\noindent{\bf 4.2 Results}
\smallskip
Firstly, the point-source models are generally a poor fit, giving
a large $\chi^2$. Using the estimated noise errors on $\chi^2$, we can
quantify the rejection of a point-source profile as  
$\sim 2\sigma$ for 6C1017+37, but $\sim 6$--$20\sigma$ for the other 9 radio galaxies. Even for the QSO, a pure point-source is rejected by $6.7\sigma$
 For all objects except 6C1256+36 and the QSO,
 there is a disk model and/or a bulge model with a
$\chi^2$ sufficiently low to indicate consistency within $2\sigma$.
In some cases, $\chi^2$ is very similar for the best-fit disk and bulge models, for other
galaxies one model may be favoured by as much as several $\sigma$.

Figure 3 shows histograms of the area of each radiogalaxy above the 8 
thresholds, together with the areas derived from the point-source model and the
best-fitting disk and bulge models. To compare the models and data more directly, we also show radial intensity profiles. Figure 4 shows
the mean $K$-band intensity in $\Delta(r)=0.5$ arcsec annuli about the centroid of each radio galaxy, together with the
 point-source profile and the best-fitting
disk and bulge models.

 Table 2 lists the apparent surface brightness (SB) of each galaxy within the central 0.5 arcsec. This will correspond only approximately
to the intrinsic properties of the galaxies, but does show some differences
between them. The mean apparent central SB is $19.07\pm 0.13$ $K$ mag $\rm arcsec^{-2}$ for the 10 6C radio galaxies, but the QSO is significantly
higher in central SB and the galaxies  6C1017+37 and 6C1129+37 are particulaly low. The morphologies of the
individual galaxies are described below.
\bigskip
\noindent {\bf 5. Morphology of Individual Radio Galaxies}  
 \smallskip
\noindent 1. 6C0822+39 appears symmetric with a profile well-fitted by the 
bulge model, which is strongly ($\sim 4\sigma$) favoured over a disk model. This galaxy appears in the $K$-band to be a normal elliptical.
\smallskip
\noindent 2. 6C0943+39 is small symmetric object, consistent with either disk or bulge at the $1\sigma$ level. On the basis of this data, it
 could be another normal elliptical.
\smallskip
\noindent 3. 6C1011+36 is small and asymmetric with a much fainter
secondary nucleus  1.7 arcsec from the 
main nucleus but clearly within the outer envelope,  suggesting that it is currently merging with a much smaller galaxy. It may also be interacting with two close companions, which produce the peak in the profile at
$r\sim 4$ arcsec. Neither a disk or a bulge profile is strongly favoured.
\smallskip
\noindent 4. 6C1017+37 is another small symmetric object, apparently the smallest of these galaxies. It is consistent with   either disk or bulge models 
at the $1\sigma$ level, and the only galaxy consistent at the $< 3\sigma$ level with
 being a point-source. The low central SB is presumably due to the object being faint and unresolved, rather than being an intrinsically low SB galaxy.  There is no excess above the fitted models at either
small or large radii, so this object could simply be a small elliptical.
\smallskip
\noindent 5. 6C1123+34 is round and symmetric, with a
 profile consistent with both the disk and bulge models.  There is another galaxy about 5.5 arcsec away but no obvious
indication that the two are interacting. This could be a normal elliptical.
\smallskip
\noindent 6. 6C1129+37 is extended and asymmetric with two intensity peaks
of similar luminosity. These lie well within the same envelope, so that PISA detects 6C1129+37 as a single galaxy even with the deblend option. The outer regions appear disturbed with some evidence of trails, especially to the north. This is the only galaxy in this sample for which the profile
significantly ($\sim 4\sigma$) favours an exponential model over a bulge model. 
It also has a low central SB (0.49 mag less than the sample mean) which, as the galaxy is clearly resolved, would
imply that the intrinsic central SB is lower than that of the other radio galaxies. 
A recent red-band WFPC2 image of this galaxy, to be 
described in detail by Best et al. (in preparation), also showed that

(i) the two peaks seen on the $K$ image 
(centroids separated by 1.9 arcsec) lie within elongated structures with tails
pointing {\it inwards} to the galaxy centre, suggesting that they are
`hotspots' associated with the outward passage of the radio jets rather than the two components of a merger. Close to the centre the jets appear to be aligned with the 137 degree position angle   
of the two peaks of radio emission (Allington-Smith 1982; Eales et al. 1997), although the two radio lobes lie much further out, being
separated by 14 arcsec. Further out from the centre, the $R$-band jets curve slightly toward a W--E
axis, causing the $K$-band peaks to be slightly misaligned  from the radio axis.

(ii) The faint extension to the north is a small spiral galaxy of
high $R$-band ($\rm \lambda_{rest}\simeq 3000\AA$) SB, which appears to be interacting with
the larger radio galaxy.  Allington-Smith et al. (1982) had previously found 6C1129+37 to be double on a ground-based $R$-band CCD image, with the two components separated north-south so presumably 
corresponding to the radio galaxy and this companion rather than the two radio jet hotspots. 

\noindent (iii) No central point source is visible at 
$\rm \lambda_{rest}\simeq 3000\AA$.
\smallskip
\noindent 7. 6C1204+35 appears symmetric, slightly favouring
($\sim 1\sigma$ level) a bulge over a disk profile.
There is no excess above the fitted bulge model at large radii. This
galaxy seems consistent with being a normal elliptical.
\smallskip
\noindent 8. 6C1217+36 is extended with a bulge profile favoured over
an exponential disk by $\sim 2\sigma$. 
 The outer regions
look slightly asymmetric, with some isophotal twist apparent on the contour plot.
 The radial profile is very close to the 
bulge model except for a small excess at $r\sim 5$ arcsec due to a much smaller companion galaxy. This could, for example, be a normal elliptical just slightly 
perturbed by a near-miss encounter, resulting in a slight asymmetry.
\smallskip
\noindent 9. 6C1256+36 is extended and double with a less luminous
secondary nucleus, 2.5 arcsec from the main nucleus. The outer regions are
 asymmetric and much more extended than in the other double galaxy (6C1011+36), with the second nucleus lying within what appears to be an inclined disk. The galaxy may also be interacing with its much smaller nearby companion (there a is hint of a bridge between the two). The galaxy is 
 not fitted well by either a disk or bulge
model, having too much area at the lowest thresholds.  However, the radial profile looks 
closer to the bulge model at $r<2$ arcsec, and the central SB is typical of this sample
 (in contrast to 6C1129+37), suggesting that the primary galaxy is an
elliptical. This is consistent with the visual impression of an elliptical merging with a large, lower SB disk galaxy.
 \smallskip
\noindent 10. 6C1257+36 is closer to a bulge profile than a disk, although only by 
$\sim 1\sigma$. The profile is well-fitted by the bulge model except
for an excess at $r\sim 4$ arcsec, due to a small companion galaxy. This 
galaxy could be another normal elliptical.
\smallskip
\noindent 11. The QSO 5C6.8 contains a very bright central nucleus, giving a central SB higher than that of any of the 10 radio galaxies, and a
very small $r_{hl}$.  However, 
the object is not a pure point source. The $K$-band image also shows the 
extended, lower surface brightness host
galaxy, which appears disturbed, suggesting a recent
merger or interaction.  The profile is close to the fitted models at $r<3$ arcsec but shows a significant excess at $r\simeq 4$ arcsec, where it is more extended and resembles the profiles of the radio galaxies.
Consequently, neither the disk or bulge models are
are a good fit, and a pure point-source model is also rejected. Two-component models and higher resolution data would be needed to investigate two-component
systems of this type, by separately estimating a point-source contribution and an $r_{hl}$ for the
underlying galaxy (see Best et al. 1997). 
\smallskip
Despite the long wavelength of observation, these galaxies show
a significant morphological diversity as well as
a wide range of half-light radius. 
Two galaxies, 6C1011+36 and 6C1256+36 appear 
to be merging doubles. The 6C1129+37 galaxy is especially unusual - it is interacting and has luminous features associated with the radio emission visible on the 
$K$-band image. We refer to this for now as a `radio jet object'. 

 Dunlop and Peacock (1993) and Ridgway and Stockton (1997) found that
many $z\sim 1$ 3CR galaxies do show an
alignment between their $K$-band isophotes and radio axes, although no
 alignment
was seen for Parkes radio galaxies an order of magnitude less radioluminous.
Our 6C sample is intermediate in radio luminosity, so might be expected to contain at least one galaxy with visible $K$-band/radio alignment.
The 6C1129+37 galaxy appears to be the only galaxy of these 10 with resolved
aligned features, but it may also be significant that 6C0943+39, 6C1017+35 and 6C1204+35 show a close alignment (within $\sim 10$ degrees) of their $K$-band position angles and radio axes (Eales et al. 1997).

The 6C1129+37 galaxy is also the only object in this sample with a $K$-band profile which appears to significantly favour
 an exponential over a bulge model. Further observations may be needed to determine whether 
it is truly a disk galaxy or an elliptical with a
profile distorted by the interaction and radio hotspots,
but the low central SB compared to the other large double galaxy (6C1256+36),
might favour a later Hubble type. 

The other seven galaxies appear at least consistent with the elliptical morphology seen for
most radio galaxies, so will
be referred to for now as `bulge/generic' galaxies.  We now consider the
relation between angular size, luminosity and morphology for these and
other radio galaxies.

  \bigskip
\noindent {\bf 6. Size vs. Absolute Magnitude Relation}
\smallskip
We estimate the absolute magnitudes of the radio galaxies in
the rest-frame Cousins $R$-band ($\rm \lambda_{cent}\simeq 6560 \AA$), by
subtracting the distance modulus  from
each observed $K$ magnitude (Table 1), and 
adding a modelled  correction to convert from the observer-frame $K$-band  to the rest-frame $R$-band.

Roche et al. (1997) describe a model for the evolution of E/S0 galaxies,  in
which star-formation begins at a high redshift, 16.5 Gyr ago, and decreases rapidly with an exponential timescale $\tau_{SFR}=0.5$ Gyr. As almost all star-formation occurs at $z>3$, the evolution is passive at the redshifts
we are concerned with here.
The star-formation history is converted into an evolving spectral
energy distribution using the 
stellar evolution models of Charlot and Bruzual (1993), and assuming a Salpeter IMF. 
Figure 5 shows the $R_{rest}-K_{obs}$ correction, as a function of redshift,  derived from this model's evolving 
spectral energy distribution, and also shows
 $R_{rest}-K_{obs}$ from much a 
bluer model in which  star-formation
begins at the same epoch but continues at a constant rate.

 At $z=1.1$, the constant SFR  model predicts much bluer observer-frame colours
than the E/S0 model, $R-K=3.5$ 
compared to $R-K=5.5$, but the $R_{rest}-K_{obs}$ 
correction differs by only 0.21 mag. 
At $z\sim 1$, many radio galaxies have $R-K$ colours close to a passively evolving model, which essentially defines the red envelope of their colour-redshift
distribution, but some of the more radioluminous (i.e. 3CR) galaxies are as much as $\sim 2$ mag bluer and would be closer to the constant SFR model
(Lilly and Longair 1984; Dunlop and Peacock 1993).
Assuming an E/S0 model should therefore give a conservative
estimate of $M_R$, accurate for the redder galaxies but an underestimate by as much as 0.21 mag for any objects which are as blue as the bluest 3CR galaxies. 

Table 3 gives our estimates of $M_R$. 
 All of these 6C  galaxies are more luminous than
the zero-redshift $L^*$ ($M_R\simeq -22.57$ for ellipticals), with a
mean $M_R$ for the radio galaxies (excluding the QSO)
of $-24.25\pm 0.18$.  
To investigate the $M_R-r_{hl}$ relation, we adopt the disk model size estimate, $r_{hl}=1.68 r_{exp}$ for all objects with a smaller minimum $\chi^2$ for the disk model (6C1011+36, 
6C1017+37, 6C1129+37, 6C1256+36) and the bulge model estimate $r_{hl}=r_{e}$
for the others. Note that adopting instead the bulge-model $r_{hl}$ for the four objects with a lower $\chi^2$ for the disk-model would increase the
 errors on their
$r_{hl}$ estimates, but would not significantly change their best-fit $r_{hl}$.
If disk galaxy is observed at an inclined angle its apparent 
area will be reduced. To estimate the intrinsic size 
of inclined galaxies we would obviously use $r_{hl}$ on the major axis rather than the smaller $r_{hl}$ derived from the apparent area.    
The observed area relative to that seen if the galaxy was seen face-on will
be approximately ($1-e$), where $e$ is the image ellipticity as measured by
PISA. We therefore correct the $r_{hl}$ estimates for inclination
by multiplying by $(1-e)^{-0.5}$. Table 3 gives the ellipticities and 
corrected $r_{hl}$ estimates, again with $\pm1\sigma$ errors, both in arcsec and
in kiloparsecs for the
assumed cosmology. The unweighted mean of the $r_{hl}$ estimates for the radio galaxies (excluding the QSO) is $6.1\pm1.3$ kpc.

 Figure 6 shows $r_{hl}$ against $M_R$ for these galaxies, with
symbols indicating morphological types, together with
 the 21 3CR galaxies from Best et al. (1997), with the fitted bulge-model radii 
converted to kpc for our assumed cosmology. These 3CR galaxies cover a wide redshift range of
$0.47<z<1.27$, with $z_{mean}=0.89$ compared to our sample's 
$z_{mean}=1.11$, so were divided into two subsamples, the 9 
galaxies at $z>0.93$, with  a mean redshift ($z_{mean}=1.10$)
 alomst identical to our 6C sample, and the 12 at $z<0.93$, with $z_{mean}=0.73$. The absolute magnitudes of the 3CR galaxies were 
estimated using the $K$-band
aperture magnitudes of Lilly and Longair (1984), extrapolated to total
magnitudes assuming bulge profiles with $r_{e}$ from Best et al. (1997), and adopting the same E/S0 model $R_{rest}-K_{obs}$ 
correction as used for the 6C galaxies.

The 3CR galaxies are on average 
 more luminous (mean $M_R=-24.88\pm 0.15$) and much 
larger (mean $r_{hl}= 17.9\pm1.5$ kpc) than the 6C galaxies. The 3CR
 galaxies at $0.93<z<1.27$ are more luminous (mean $M_R=-25.30\pm 0.16$) than
those at $0.47<z<0.93$ (mean $M_R=-24.56\pm 0.19$, but not significantly different in size (mean $r_{hl}=17.5\pm 1.7$ kpc compared to 
$18.2\pm 2.4$ kpc). The 3CR galaxies at $0.93<z<1.27$ are on average $1.05\pm 0.24$ mag more luminous in the $R$-band, and $\sim 0.46$ dex larger, than our 6C
galaxies at the same mean redshift (note that
this result will not depend significantly on the
assumed $q_0$). This magnitude difference is greater than the 0.7 mag
reported by Eales et al. (1997) for a $z>0.6$ sample, probably as a result of our exclusion of the less evolved 3CR galaxies at $0.6<z<0.93$ and our use of total
rather than metric magnitudes (giving a greater difference in luminosity
between small and large galaxies).

To compare these galaxies with those at lower redshift, Figure 6 shows  local $r_{hl}-M_R$ relations, converted from the $r_{hl}-M_{B}$ relations used by e.g.
Roche et al. (1997) to the $R$-band assuming the $z=0$ colour of $B-R=1.44$ for ellipticals (as given by the model)
and $B-R=1.0$ for spirals.

\noindent For ellipticals, the Bingelli, Sandage and Tarenghi
(1984) relation becomes
$${\rm log}(r_{hl}/{\rm kpc})=-0.3(M_R +20.01)$$
for $M_R<-21.44$, and
$${\rm log}(r_{hl}/{\rm kpc})=-0.1(M_R +17.14)$$
for $M_R>-21.44$. 

\noindent For spirals, the Freeman (1970) surface brightness becomes,
$${\rm log}(r_{hl}/{\rm kpc})=-0.2M_R -3.42$$

Figure 6 also shows an observed sample of low redshift radio galaxies, from
Owen and Laing (1989), who estimated
absolute magnitudes in the Cousins $R$ band and bulge-model-fitted
 radii for $z<0.2$ radio galaxies of a number of types. We plot their Classical Double FRII (Fanaroff and Riley 1974) radio galaxies, the type most appropriate for comparison with the higher redshift samples,  with sizes and magnitudes converted to $H_0=50$ km $\rm s^{-1} Mpc^{-1}$ and
their isophotal absolute magnitudes extrapolated to total magnitudes assuming bulge profiles
with their fitted effective and isophotal radii. These 24 low redshift
($z_{mean}=0.11$) radio galaxies are less luminous  
(mean $M_R=-23.37\pm 0.12$) than those at higher redshift, and have a 
very wide range of sizes with a mean  $r_{hl}$ of $12.5\pm  2.3$ kpc, intermediate between the 6C and 3CR galaxies.

 The QSO is displaced far from the all other sources on this plot, on account of its central point-source. 
As we cannot measure the true galaxy size with this data, this
object must be  excluded from our discussion of $r_{hl}-M_R$
relations. 

The 6C galaxies are clearly shifted relative to the $r_{hl}-M_R$ relation of local ellipticals, in the direction of a higher intrinsic SB.
 As noted by Best et al. (1997), the
  3CR objects also tend to be enhanced in SB
relative to local ellipticals. Note that 
estimates of the intrinsic SB will be independent of the assumed $q_0$ and $H_0$. The $r_{hl}-M_R$ relations of 
both the $z\sim 0$ and  $z\sim 1$ radio galaxies do not appear to  follow the $\Delta({\rm log}~r_{hl})=-0.2\Delta (M)$ slope of 
constant SB typical of spirals -- they are noticeably steeper and more consistent with the $\Delta({\rm log}~r_{hl})=-0.3\Delta (M)$ 
slope of the giant ellipticals.
Hence to estimate the surface brightness evolution $\Delta R$  relative to local E/S0 galaxies of the same size, we fit the function  
$${\rm log}(r_{hl}/{\rm kpc})=-0.3(M_R +\Delta R + 20.01)$$
For the
 7 6C galaxies we classed as bulge/generic types, $\chi^2$ is minimized for  
$\Delta R=1.79\pm 0.22$ mag, with a $\chi^2$ of only 6.8 indicating that all 7
are consistent with this single $r_{hl}-M_R$ relation. 
One of double objects (6C1011+36) lies on the same
relation. It appears to be undergoing just a minor merger, of a large elliptical with a much less massive dwarf galaxy, and this might not greatly affect the total size or luminosity.
 However, the
other double object (6C1256+36) and the `radio jet galaxy' 6C1129+37 are of 
lower mean SB, corresponding to
$\Delta R=1.01^{+0.17}_{-0.13}$ mag and $\Delta R=1.05^{+0.07}_{-0.17}$ mag respectively. 
The image of the 6C1256+36 galaxy was suggestive of
an elliptical merging with a large, low surface brightness 
disk galaxy. This
would obviously decrease the mean SB of the combined 
object and give it a more disk-like profile, while accounting for the 
 central SB remaining high. In contrast, the radio jet object 6C1129+37 
has a intrinsically low central SB as well as a low mean SB.

An error-weighted fit to all 10 6C objects gives $\Delta R=1.26\pm0.18$ mag, 
but with a high $\chi^2$ of 62.16 indicating that the galaxies are not all 
consistent with a single $r_{hl}-M_R$ relation -- there is a significant 
difference between the two relatively low mean SB objects and the
relation defined by the other 8. 
An unweighted fit of the same relation to the 24 low redshift 
FRII galaxies gives 
$\Delta R=0.12\pm 0.19$ mag, consistent with unevolved ellipticals. 
An unweighted fit to the 21 3CR galaxies gives evolution of 
$\Delta R=0.80\pm 0.15$ mag, rather less  than for our sample. However, 
if we separately consider the lower redshift ($0.47<z<0.92$) and higher redshift ($0.92<z<1.27$) 3CR subsamples, the former gives $\Delta R=0.51\pm 0.18$ and
the latter $\Delta R= 1.20\pm 0.18$. There appears to be surface brightness evolution
of $0.69\pm 0.25$ mag  
between the respective mean redshifts of 0.73 and 1.10, with the higher redshift
subsample being much closer in size-luminosity relation to the 6C galaxies at $z_{mean}=1.11$. We discuss the implications in Section 7 below.
\bigskip
\noindent {\bf 7. Discussion} 
\smallskip
\noindent{\bf 7.1 Surface Brightness Evolution}
\smallskip
Using ground-based data, we have carried out a preliminary investigation of the $K$-band angular size and morphology of 6C radio galaxies at $1<z<1.4$. Our first result is that the
6C galaxies typically  have smaller
$K$-band 
half-light radii than 3CR radio galaxies at similar redshifts. This appears to
exclude the hypothesis that 3CR and 6C sources at $z\sim 1$ 
have identical host galaxies and the higher $K$-band luminosity of the 3CR sources (e.g. Eales et al. 1997) is
due solely to their more powerful active nuclei. If this was the case, the 3CR 
galaxies would have {\it smaller} half-light radii than their 6C counterparts, due to the greater dominance of the
central point-sources. The difference in $K$ luminosity 
appears to be  due primarily to the stellar components, which in the 3CR galaxies are considerably larger and presumably more massive.

We estimate that the $z_{mean}=1.11$ 6C galaxies show a mean surface brightness enhancement of
 $1.26\pm 0.18$ mag in the rest-frame Cousins $R$-band, relative to the  $r_{hl}-M_R$ relation of local ellipticals. This degree of evolution is consistent with the $1.20\pm 0.18$ mag we estimate for 
 a subsample of more radioluminous 3CR galaxies at the same mean redshift.
However, the 6C galaxies which most resembled normal
 ellipticals appeared to show greater SB evolution
of $1.79\pm 0.22$ mag. One possible explanation for this higher 
estimate is that some radio galaxies contain central point sources, which with the limited resolution of our data are not distinguished from the underlying 
galaxy but do cause an underestimate of $r_{hl}$. This would cause the SB evolution to be overestimated, and if the effect on $r_{hl}$ is 
very large might even account for the difference in the mean $r_{hl}$ of
the 3CR and 6C galaxies. 
 To estimate the likely effect of point-source contamination, we consider a 
bulge-profile model with $r_{e}=1.8$ arcsec -- the mean size of the Best et al. (1997) 3CR galaxies -- to which is added a central point-source contributing from
$0\%$ to $100\%$ of the total flux. These models were convolved with seeing profiles and
normalized to have $K=17.4$ with sky noise typical of the real Redeye data, and then
analysed in the same way as the real radio galaxy images.

 Figure 7 shows the
estimated bulge-model $r_{e}$ as a function of the point-source
contribution to the total flux. A point-source component of $10\%$ reduces the fitted $r_{e}$ by
$\sim 25\%$, so could increase the estimated $\Delta R$ by 0.42 mag. 
This would be sufficient to account for much of the difference in the estimated
$|Delta R$ between the 
bulge/generic 6C galaxies and the 3CR galaxies. However, to reduce the
fitted $r_{hl}$ from 1.8 arcsec to 0.55 arcsec, the mean size of the 6C galaxies, would
require a much larger point-source contribution of $\geq 30\%$. It is extremely
unlikely that such luminous point-sources are typical of the 6C radio galaxies.
Although we are at present only able to exclude a strong central point-source in the case of 6C1129+37 (for which we have WFPC2 data), only 2/21 of the 3CR galaxies of Best et al. (1997) had profiles consistent with any point-source 
component of $\geq 10\%$ in the observed $K$-band, and if anything the less radioluminous 6C galaxies would be expected to show less nuclear activity.

Hence it seems that central point-sources would not
make more than a small contribution to the difference in  
$r_{hl}$ between 6C and 3CR galaxies, but they might still be present in the bulge/generic galaxies at the 
$<10\%$ level and be sufficient to cause an
overestimate of the surface brightness evolution 
by up to $\sim 0.42$ mag. This would still leave at least $\sim 1.37$ mag
of genuine evolution. This is still less than the surface brightness enhancement of $z\sim 3$ galaxies on the Hubble Deep Field (e.g. Jones and Disney
1997; Roche et al. 1997b), so there is no doubt that such a degree of SB evolution is possible for large galaxies at $z>1$. 
 
In constrast, the SB of FRII radio galaxies at $z_{mean}=0.11$ showed
only $0.12\pm 0.19$ mag of evolution relative to the local elliptical $r_{hl}-M_R$
relation, so these objects are consistent with being unevolved ellipticals or 
with
the 0.12 mag of passive evolution predicted by our model at their mean redshift.
  This confirms that the assumed $r_{hl}-M_R$ relation is appropriate for
powerful radio galaxies at low redshifts as well as for normal ellipticals,
 and hence that
the $r_{hl}-M_R$ relation of FRII radio galaxies in particular does shift with 
increasing redshift. As would be expected, a $z_{mean}=0.73$ subsample of 3CR galaxies showed
intermediate evolution, estimated as $0.51\pm 0.18$ mag. 

This progressive
shift of the $r_{hl}-M_R$ relation with increasing redshift is similar to
that seen for normal radioquiet ellipticals over a similar redshift range.
Schade et al. (1997) found the rest-frame blue-band surface brightness evolution of a mixture of field and cluster ellipticals to be best-fitted by
$0.78z$ mag for $q_0=0.5$, or $\sim 0.92z$ mag for a low $q_0$.
 This is
consistent with the passively evolving E/S0 model described in Section 6, which predicts $\Delta(M_B)=-0.91$ mag at
$z=1$. In the rest-frame Cousins red-band, the model predicts slightly less
evolution of $\Delta(M_R)=-0.71$ mag, with 0.63 mag at $z=0.73$, consistent with observations for
the 3CR galaxies, and 0.75 mag at $z=1.11$, a little less than observed. 
 Hence there may be at least marginal indications that some $z>1$
radio galaxies in both the 3CR and 6C catalogs undergo more optical evolution 
than expected from passive evolution. This might be explained as the more rapid fading of those components
of the radio galaxy luminosity
 not produced by old stellar populations, e.g. young stars
formed along the radio jets (see Best et al. 1996), and diffuse scattered light from the central AGN. Note also that a steeper IMF could
increase the rate of passive luminosity evolution.
\smallskip
\noindent{\bf 7.2 Implications for Radio Galaxy Evolution}
\smallskip

The simplest evolutionary scenario is that surface brightness changes but
$r_{hl}$ does not. Although there is probably some increase with time in the mean size of galaxies, angular sizes from WFPC2 data suggest
that, for ellipticals, this is a small effect out to $z\sim 1$ (e.g. Im et al. 
1996; Roche et al. 1997). 
 Assuming no evolution
in $r_{hl}$, the estimated radii of the $z\sim 1.1$ 6C galaxies would on the basis of the assumed local $r_{hl}-M_R$
relation  indicate their local counterparts to
have a mean absolute magnitude $M_R=-22.6$, i.e. about $L^*$, and therefore suggest that 
many evolve into $L\sim L^*$ ellipticals. Most of the $z<0.2$ FRII radio galaxies in the Owen and Laing (1989) sample are
ellipticals of modest size and $L\sim L^*$ luminosity, although
quite high radio luminosities 
of $L(1400 $MHz$)\simeq 10^{26.3}$ $\rm WHz^{-1}$. 
It is tempting to identify these as the low redshift counterparts of our 6C galaxies. Their typical radio luminosity corresponds  
to $L(151 $MHz$)\simeq 10^{27.2}$ $\rm WHz^{-1}$ for the average SED of luminous radio galaxies ($\alpha\simeq - 0.94$), so this would require a 
factor of $\sim 25$ decrease in radio luminosity since $z\sim 1.1$. This is  consistent with the 6C radio luminosity function undergoing the same $\tau=3.3$ Gyr rate of evolution as estimated (Padovani and Urry 1992) for that of 3CR galaxies, but of course if our identification is correct the evolution of the radio luminosity function  must be due to the evolution of individual objects rather than simply
a change in the number of galaxies which become powerful radio sources.

Owen and Laing (1989) and Owen and White (1991) found that  the  classical double FRII radio galaxies at $z<0.2$  were typically $\sim 10^{1.6}$ higher in 
radio luminosity ($L_{rad}$) than FRI galaxies
of similar optical luminosity, with the dividing line between the classes following a $L_{rad}\propto L_{opt}^2$ relation (Ledlow and Owen 1996). 
As previously reported by e.g. Longair and Seldner (1979) and  Prestage and Peacock (1988, 1989), the FRI radio galaxies tended to lie within clusters (on average Abell class 0), while nearby FRIIs were usually in field 
environments.
 However, Yates, Miller and Peacock (1989) and Hill and Lilly (1991) found 
that powerful radio galaxies at higher redshifts
 ($z\sim 0.5$) were as likely to lie in Abell class 0-1 clusters as 
in the field, suggesting a more rapid $L_{rad}$ evolution for the 
 sources in clusters.

If any of these 6C galaxies lie in rich clusters, their low redshift counterparts might instead be FRI galaxies, like the smaller of those 
in the Owen and White (1991) sample, with much lower 
present-day radio luminosities of $L(1400 $MHz$)\simeq 10^{24.7}$
 $\rm WHz^{-1}$. This corresponds to 
$L(151 $MHz$)\simeq 10^{25.6}$ $\rm WHz^{-1}$, a factor of $\sim 1000$ down
on the $L_{rad}$ of our 6C galaxies. Yee and Ellingson
(1993) estimate that the optical luminosity evolution of AGN in rich
cluster environments (Abell class $\sim 1$) is as rapid as $\tau\simeq1.0\pm 0.2$ Gyr out to $z\sim 0.6$, but at higher redshifts 
must be similar to other AGN. If this form of evolution is
paralleled by the evolution of $L_{rad}$ it would predict a decrease in
$L_{rad}$ of a factor of $\sim 1000$ since $z\sim 1.1$, exactly the amount
required.

We estimate that the mean $M_R$ of 3CR and 6C galaxies 
at $z\sim 1.1$ differs by $1.05\pm 0.24$ mag. The mean radius of the 3CR galaxies is much larger and with no size evolution would correspond to low-redshift counterparts with $M_R=-24.2$, several $L^*$.
On this basis we suggest that the difference in 3CR and 6C luminosity 
at $\lambda_{rest}\sim 1\mu\rm m $ is due to a difference in the stellar masses of the the host galaxies.  As the radio luminosities differ by a factor $\sim 6$, this suggests a correlation 
$L_{rad}\propto M_{stellar}^2$.

A possible scenario explaining these observations is that;

(i) At $z\geq 1$, the maximum radio luminosity is strongly correlated with the stellar 
mass of the host galaxy, approximately as $L_{rad}\propto M^2$. This relation could physically be related to the higher pressure of  
surrounding gas, and perhaps the  more
massive central black hole, in a more massive galaxy (see e.g. Eales 1992; Rawlings and Saunders 1991).
 At $z>1$ many of the 3CR and 6C galaxies may be observed while close to this upper limit, with the 3CR galaxies typically 
being $\sim 2.5$ times more massive and hence $\sim 6$ times more radioluminous.

(ii) High redshift radio galaxies exist in a mixture of field and cluster environments. The peak $L_{rad}$ attained during subsequent bursts of radio emission decreases more rapidly for radio galaxies in a rich cluster
environment (Yee and Ellingson 1993).
 By $z\sim 0$, the cumulative effect of this environmental 
dependence amounts to a factor of $\sim 10^{1.6}$ in $L_{rad}$ between Abell class 0 environments 
and the field, obscuring the $L_{rad}\propto M^2$ correlation seen at $z\geq 1$. However, as cluster environment also influences
the evolution of the radio morphology, a $L_{rad}\propto M^2$
correlation reappears when nearby radio galaxies are separated 
into FRI and FRII types (Ledlow and Owen 1996). 
 
\medskip
At any rate the high surface brightness and relatively small size of the 6C galaxies seems to argue that their 
evolution, at both 
 optical and radio wavelengths, is no less rapid than that of the 3CR
galaxies. Although the $K-z$ relation of 6C galaxies may have appeared consistent with no evolution (Eales et al. 1997), their $r_{hl}-M_R$
relation clearly is not. Of course,
the results of this paper, being based on a small dataset of marginally adequate depth
and resolution, can only be regarded as a preliminary. We intend to investigate larger numbers of $z\sim 1$ radio galaxies, using NICMOS and ground-based adaptive optics to better constrain half-light radius and morphology,
and to separate out bulge, disk, point-source and
radio jet components.  
\bigskip
{\bf 8. Summary of Main Conclusions}
\smallskip
(i) A sample of 6C radio galaxies at $1<z<1.4$, observed in the $K$-band, showed significant morpological diversity. Two appeared to be undergoing mergers, and another
possessed two bright peaks aligned with the radio jets and may be closer to an 
exponential than a bulge profile, but the other seven were more consistent with normal elliptical morphologies in the $K$-band. 
\smallskip
(ii)  We estimated a mean half-light radius of $6.1\pm 1.3$ kpc
 for the 6C galaxies, which is similar to that of 
present-day $L\sim L^*$ ellipticals and
significantly smaller than the 3CR galaxies studied by Best et al. (1997). 
The difference between the measured sizes of 6C and 3CR galaxies 
cannot be accounted for by point-source contamination,
 unless at the $\geq 30\%$ level, which seems implausible on the basis of the lack of point-sources in the 3CR galaxies. Hence the difference in 
$\lambda_{rest}\sim 1\mu \rm m$  luminosity between 3CR  and 6C galaxies at 
$z\sim 1.1$  must results from a difference in the sizes
 of the host galaxies, and not solely from
a difference in the power of the central nuclei.
\smallskip
(iii) The size-luminosity relation of the 6C galaxies at $z\sim 1.1$ is  offset from that of either local ellipticals or FRII radiogalaxies at $z<0.2$, indicating 1.0--1.8 magnitudes of surface brightness evolution
in the
 rest-frame $R$-band. This is similar to the 1.20 mag 
of surface brightness evolution seen for 3CR galaxies at the same redshift,
 suggesting that the 6C galaxies are undergoing a similar amount of evolution at optical wavelengths. 
\smallskip
(iv) The sizes of the 6C galaxies are 
consistent with their low-redshift
counterparts being FRII radio galaxies like those studied by Owen and Laing
(1989), a factor of $\sim 25$ lower in radio luminosity, suggesting
radio luminosity evolution similar to the Padovani and Urry (1992) estimate for 3CR galaxies.
Any 6C galaxies which lie in clusters may undergo a much greater reduction
in radio luminosity to become FRI objects.
\bigskip
\noindent{\bf Acknowledgements}
\smallskip
Nathan Roche acknowledges the support of a PPARC research associateship, and
some useful discussions with Kavan Ratnatunga on the analysis of faint galaxy
images. We thank Garrett Cotter for help with the observations.
\bigskip 
\noindent {\bf References}
\bigskip
\hangindent=2pc \hangafter=1 Allington-Smith,~J.R., 1982.
 {\it Mon. Not. R. astr. Soc.\/}, {\bf 199}, 611.

\hangindent=2pc \hangafter=1 Allington-Smith,~J.R., Perryman,~M., Longair,~M.S.,
Gunn,~J.E. and Westphal,~J.A., 1982.
 {\it Mon. Not. R. astr. Soc.\/}, {\bf 201}, 331.

\hangindent=2pc \hangafter=1 Best,~P.N., Longair,~M.S. and R\"ottgering,~H.,
1996. {\it Mon. Not. R. astr. Soc.\/}, {\bf 280}, L9.

\hangindent=2pc \hangafter=1 Best,~P.N., Longair,~M.S. and R\"ottgering,~H.,
1997. Preprint.

\hangindent=2pc \hangafter=1 Binggeli,~B., Sandage,~A. and Tarenghi,~M. 1984.
{\it Astron. J.\/}, {\bf 89}, 64.

\hangindent=2pc \hangafter=1 Bruzual,~G. and Charlot,~S., 1993.  
{\it Astrophys. J.\/}, {\bf 405}, 538.

\hangindent=2pc \hangafter=1 Dunlop,~J.S. and Peacock,~J.S.,
1993. {\it Mon. Not. R. astr. Soc.\/}, {\bf 263}, 936.

\hangindent=2pc \hangafter=1 Eales,~S., 1982. {\it Astrophys. J.\/}, {\bf 397}, 49.

\hangindent=2pc \hangafter=1 Eales,~S., 1985. {\it Mon. Not. R. astr. Soc.\/}, {\bf 217}, 149.

\hangindent=2pc \hangafter=1 Eales,~S. and Rawlings,~S., 1996.
{\it Astrophys. J.\/},  {\bf 367}, 1. 

\hangindent=2pc \hangafter=1 Eales,~S., Rawlings,~S., Law-Green,~D., 
Cotter,~G. and Lacy,~M., 1997. {\it Mon. Not. R. astr. Soc.\/}, in press.

\hangindent=2pc \hangafter=1 Fanaroff,~B.L. and Riley,~J.M., 1974.
{\it Mon. Not. R. astr. Soc.\/}, {\bf 167}, 31p.

\hangindent=2pc \hangafter=1 Freeman,~K., 1970.
{\it Astrophys. J.\/}, {\bf 160}, 811.

\hangindent=2pc \hangafter=1 Hill,~G.J. and Lilly,~S.J., 1991.
{\it Astrophys. J.\/},  {\bf 367}, 1. 

\hangindent=2pc \hangafter=1 Im,~M., Griffiths,~R.E., Ratnatunga,~K.U.,
and Sarajedini,~V.L., 1996.  {\it Astrophys. J.\/}, {\bf 461}, L79.

\hangindent=2pc \hangafter=1 Jones,~B. and Disney,~M., 1996. {\it HST and the
High Redshift Universe}, eds. N.R. Tanvir, A. Ar\'agon-Salamanca and J.V.
Wall, 37th Herstmonceux Conference, Cambridge.

\hangindent=2pc \hangafter=1 Laing,~R.A., Riley,~J.M. and Longair,~M.S., 1983. {\it Mon. Not. R. astr. Soc.\/}, {\bf 204}, 151.

\hangindent=2pc \hangafter=1 Ledlow,~M. and Owen,~F.N., 1996.
 {\it Astron J.\/}, {\bf 112}, 9.

\hangindent=2pc \hangafter=1 Leyshon,~G. and Eales,~S., 1997. 
{\it Mon. Not. R. astr. Soc.\/}, submitted.

\hangindent=2pc \hangafter=1 Lilly,~S.J. and Longair,~M.S., 1984.
{\it Mon. Not. R. astr. Soc.\/}, {\bf 211}, 833.

\hangindent=2pc \hangafter=1 Lilly,~S.J., MacLean~I.S. and Longair,~M.S.,
1984.  {\it Mon. Not. R. astr. Soc.\/}, {\bf 209}, 401.

\hangindent=2pc \hangafter=1 Longair,~M.S. and Seldner,~M.,
1979. {\it Mon. Not. R. astr. Soc.\/}, {\bf 189}, 433.

\hangindent=2pc \hangafter=1 Owen,~F.N. and Laing,~R.A., 1989.
 {\it Mon. Not. R. astr. Soc.\/}, {\bf 238}, 357.

\hangindent=2pc \hangafter=1 Owen,~F.N. and White,~R.A., 1991.
 {\it Mon. Not. R. astr. Soc.\/}, {\bf 249}, 164.

\hangindent=2pc \hangafter=1 Padovani,~P. and Urry,~C.M., 1992.
 {\it Astrophys. J.\/}, {\bf 387}, 449.

\hangindent=2pc \hangafter=1 Prestage,~R. and Peacock,~J.A., 1988. {\it Mon. Not. R. astr. Soc.\/}, {\bf 230}, 131.

\hangindent=2pc \hangafter=1 Prestage,~R. and Peacock,~J.A., 1989. {\it Mon. Not. R. astr. Soc.\/}, {\bf 236}, 959.

\hangindent=2pc \hangafter=1 Rawlings,~S., et al., 1997. In preparation.

\hangindent=2pc \hangafter=1 Rawlings,~S. and Saunders,~R., 1991.
 {\it Nature}, {\bf 349}, 138.

\hangindent=2pc \hangafter=1 Ridgway,~S.E. and Stockton,~A., 
1997. {\it Astron. J.\/},  in press.

\hangindent=2pc \hangafter=1 Rigler,~M.A., Lilly,~S.J., Stockton,~A., 
Hammer,~F., LeFevre,~O.N., 1992. {\it Astrophys. J.\/},  {\bf 385}, 61. 

\hangindent=2pc \hangafter=1 Rigler,~M.A. and Lilly,~S.J., 1994.
 {\it Astrophys. J.\/},  {\bf 427}, L79.

\hangindent=2pc \hangafter=1 Roche,~N., Ratnatunga,~K., Griffiths,~R.E.,
and Im,~M., 1997.
 {\it Mon. Not. R. astr. Soc.\/}, submitted.

\hangindent=2pc \hangafter=1 Schade,~D., Barrientos,~L. and L\'opez-Cruz,~O.,
1997. {\it Astrophys. J.\/}, {\bf 477}, L17.

\hangindent=2pc \hangafter=1 Tadhunter,~C.N., Scarrott,~S.M., Draper,~P. and Rolph,~C., 1992. {\it Mon. Not. R. astr. Soc.\/}, {\bf 256}, 53P.

\hangindent=2pc \hangafter=1 Wainscoat,~R.J., and Cowie,~L., 1992.
{\it Astron. J.\/}, {\bf 103}, 323.

\hangindent=2pc \hangafter=1 Yates,~M.G., Miller,~L. and Peacock,~J.A.,
1989. {\it Mon. Not. R. astr. Soc.\/}, {\bf 240}, 129.

\hangindent=2pc \hangafter=1 Yee,~H.K.C. and Ellingson,~E., 1993.
{\it Astrophys. J.\/},  {\bf 411}, 43.

\bigskip
\noindent {\bf Table Captions}
\bigskip
\noindent{\bf Table 1.} The co-ordinates of the 11 radio galaxies, their $K$-band magnitudes as measured from this data, their spectroscopic redshifts, and the detection thresholds of the Redeye images.
\medskip
\noindent{\bf Table 2.} A comparison of the areas of each radio
galaxy above a series of 8 thresholds with those given by model profiles; (i) the $\chi^2$ for
a point-source model; (ii) the $\chi^2$ and exponential scalelength (with
$\pm 1\sigma$ errors) of the best-fitting disk-profile model; 
(iii) the $\chi^2$ and effective radius (with
$\pm 1\sigma$ errors) of the best-fitting bulge-profile model.
\medskip
\noindent{\bf Table 3.} Estimated absolute magnitude in the rest-frame 
$K$-band, the ellipticity of the radiogalaxy images as determined by PISA, and
the half-light radius of the best-fitting model (with $\pm 1\sigma$
errors), corrected for inclination.
\bigskip
\noindent{\bf Figure Captions}
\bigskip
\noindent{\bf Figure 1.} Grey-scale plots of $15\times 15$ arcsec areas
of the $K^{\prime}$-band Redeye images centred on each radio galaxy. North is at the top, east at the left. Top row (left to right): 6C0822+39, 6C0943+39,
 6C1011+36. Second row: 6C1017+37, 6C1123+34, 6C1129+37.
Third row: 6C1204+35, 6C1217+36, 6C1256+36. Bottom row, 6C1257+36, 5C6.8.
\medskip
\noindent{\bf Figure 2.} Contour plots of the images on Figure 1.
\medskip
\noindent{\bf Figure 3.} Histograms of the area (in $0.5\times 0.5$ arcsec pixels) of each radio galaxy image above a series of 8 intensity
thresholds defined by
PISA (the lowest being the detection threshold given in Table 1). The plots also
show the areas measured above the same thresholds for a point-source (long-dashed) model,
 the best-fitting disk
model (dotted), and the best-fitting bulge model (short dash), when convolved with the seeing profile and normalized to the same total intensity as the galaxies.
\medskip
\noindent{\bf Figure 4.} Radial intensity profiles of the radio galaxies 
on the Redeye images, shown as the
mean surface brightness in  $K$ mag $\rm arcsec^{-2}$ (with $\pm 1\sigma$ errors) in annuli of 0.5 arcsec width around the image centroid, together with a point-source (long-dashed) model,
 the best-fitting disk
model (dotted), and the best-fitting bulge model (short dash), all convolved with the seeing profile and normalized to the same total intensity as the galaxies.
\medskip
\noindent{\bf Figure 5.} Modelled magnitude correction between the  observer-frame $K$-band and rest-frame Cousins $R$-band, as a function of redshift, for a passively evolving E/S0 galaxy model 
and a model with a constant star-formation rate.
\medskip
\noindent{\bf Figure 6.} Estimated half-light radius (in kpc) against
rest-frame $R$-band absolute magnitude for the 6C radio galaxies, with symbols
indicating morphology, together with the 3CR
radio galaxies from Best et al. (1997), the low-redshift FRII galaxies from  Owen and Laing (1989), and assumed
$r_{hl}-M_R$ relations for local ellipticals (dashed line) 
and spirals (dotted line).
\medskip
\noindent{\bf Figure 7.} Bulge-model half-light radius ($r_{e}$) measured 
as described in Section 4
for a simulated $r_{e}=1.8$ arcsec bulge-profile galaxy 
to which is added an increasingly luminous central point-source.
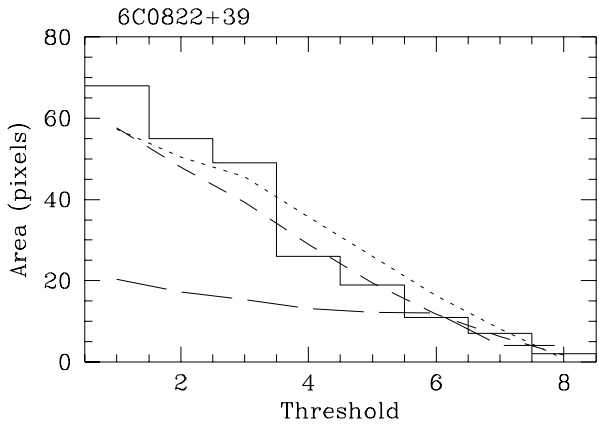
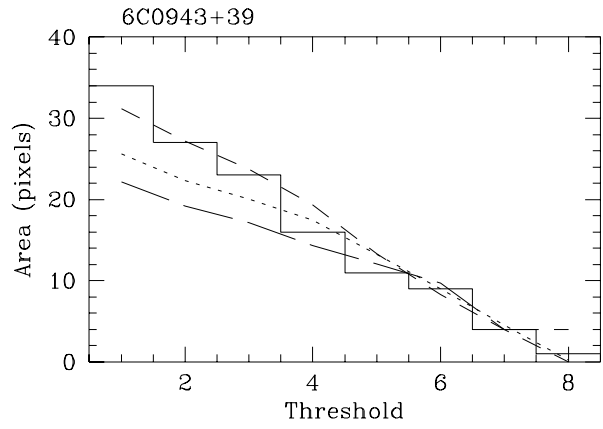
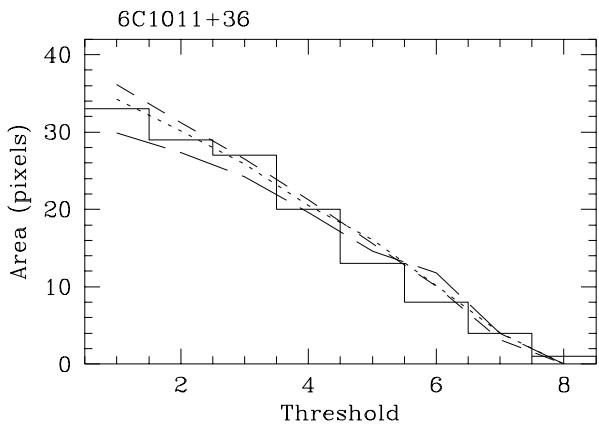
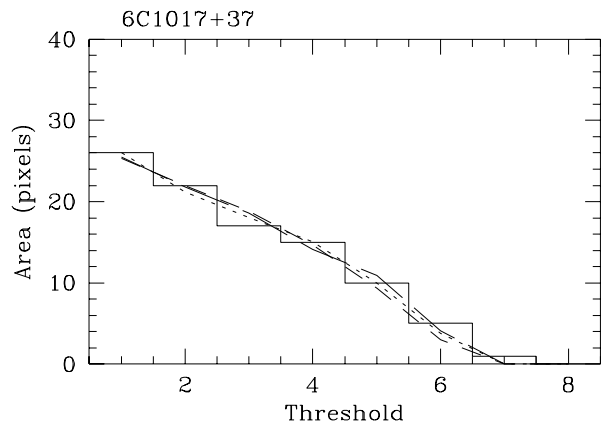
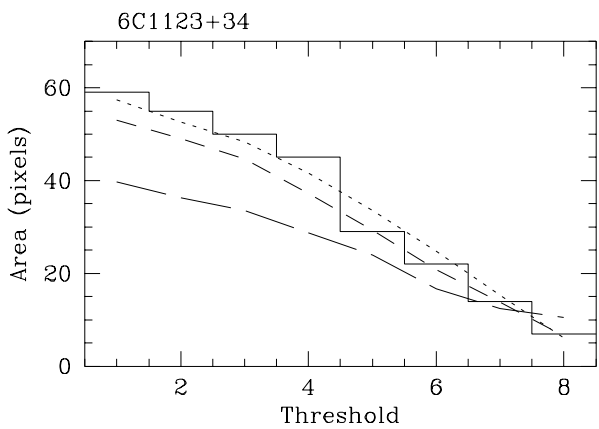
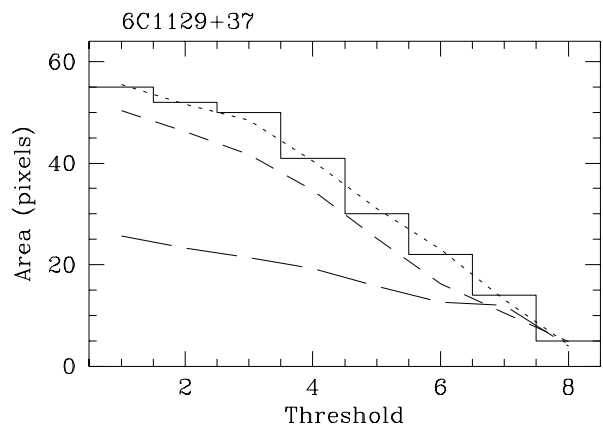
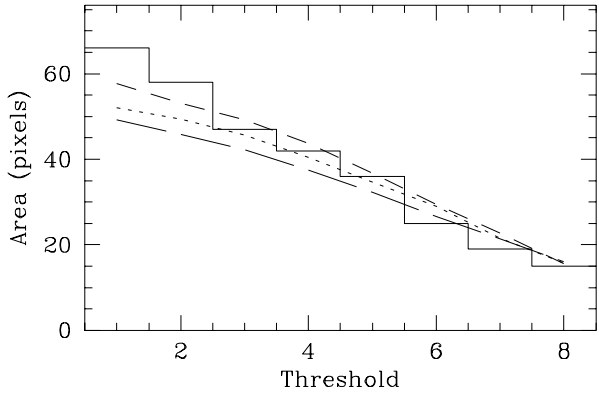
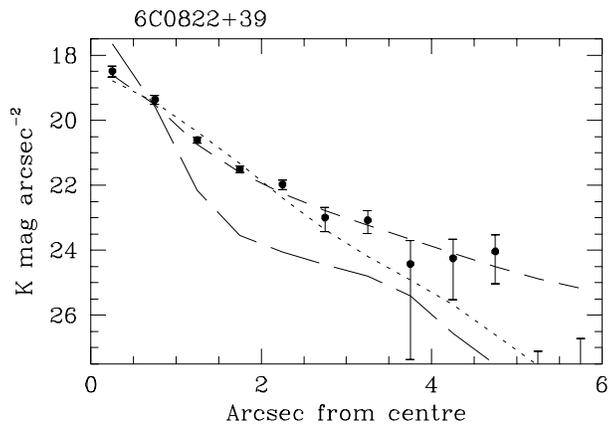
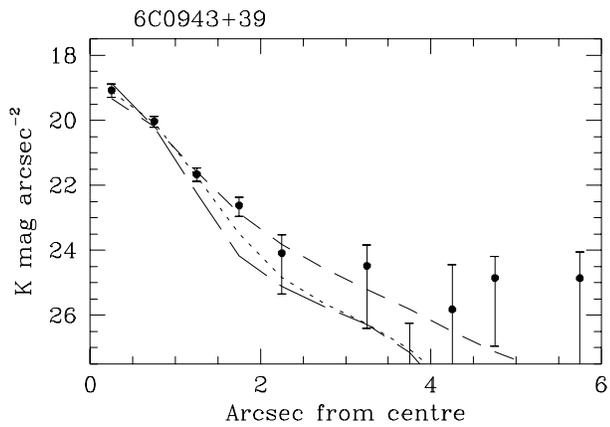
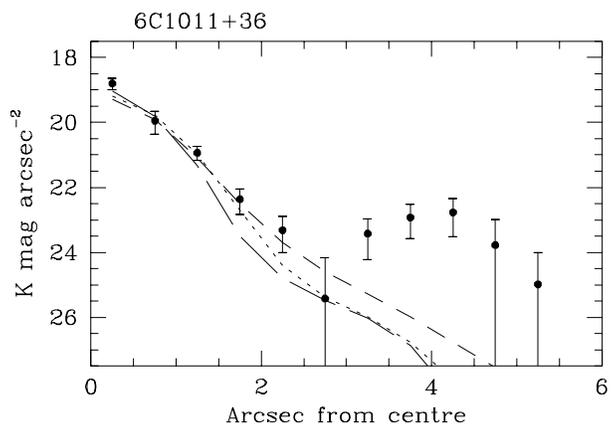
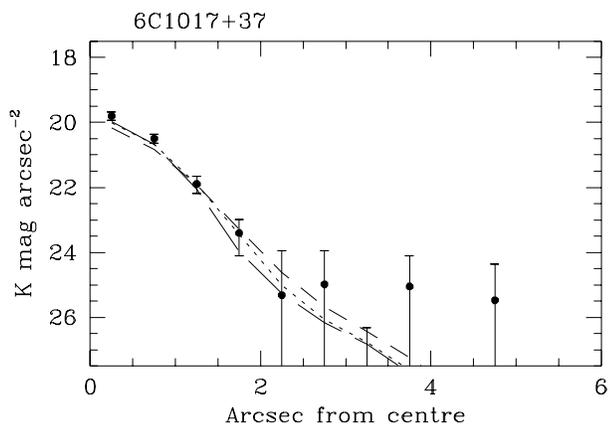
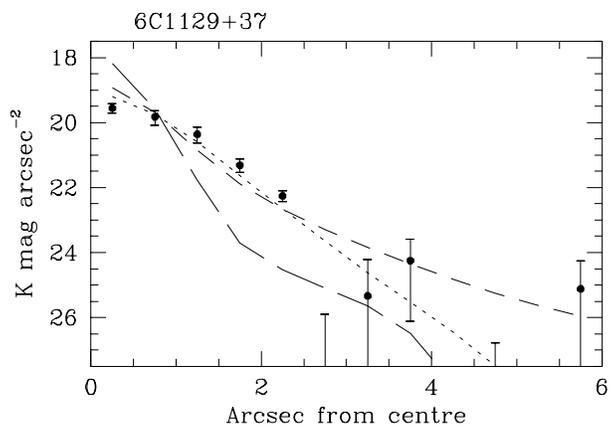
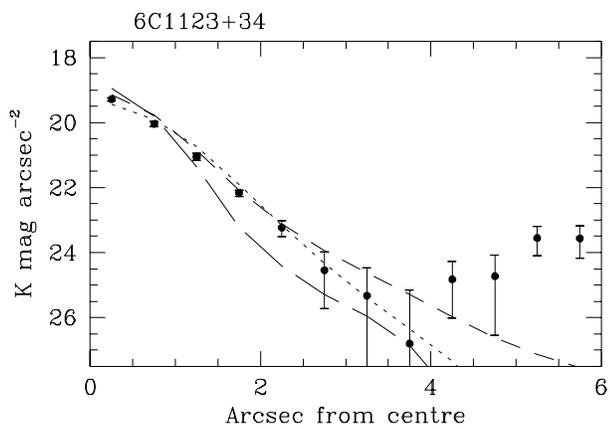
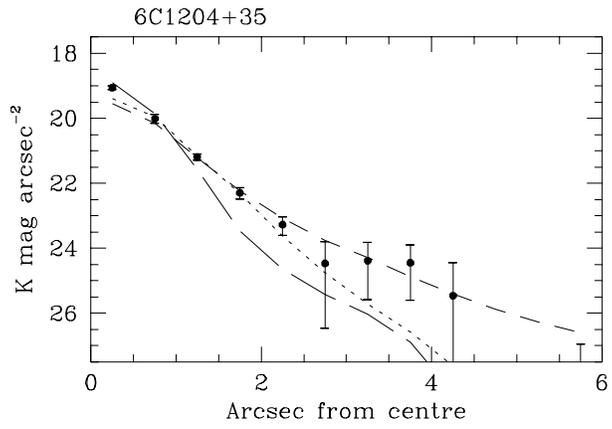
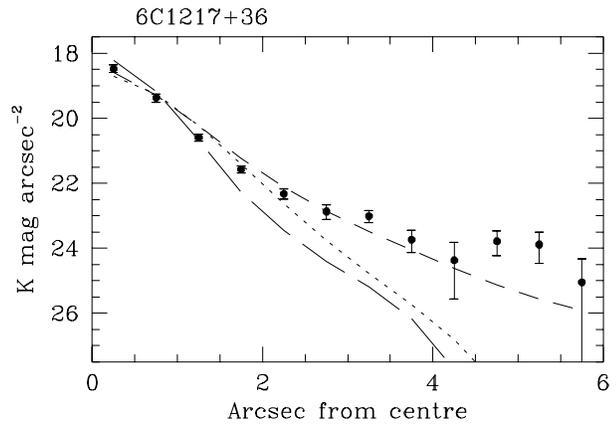
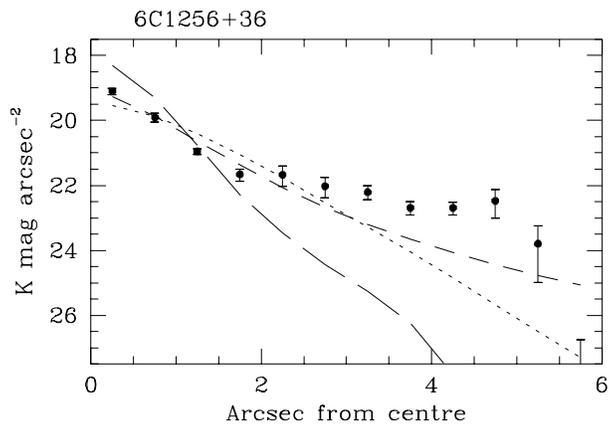
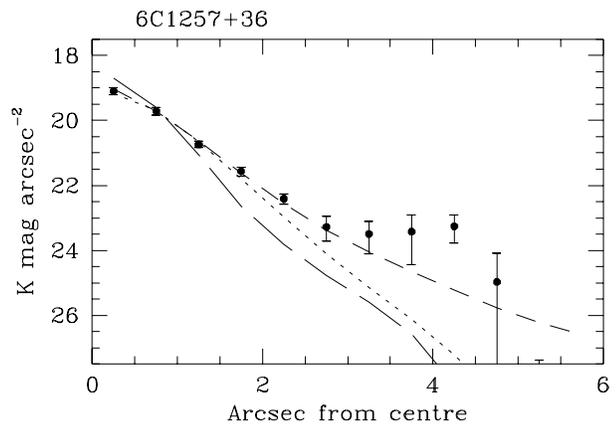
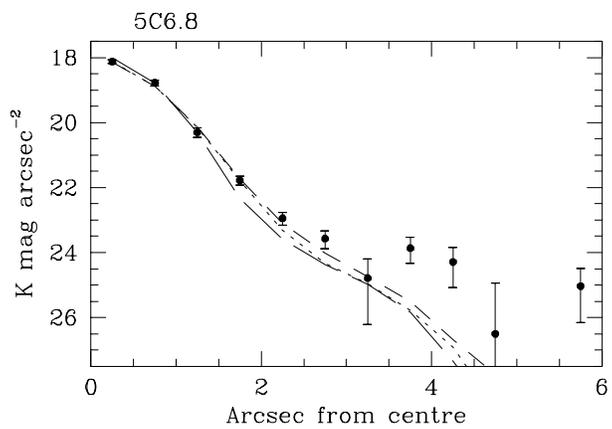
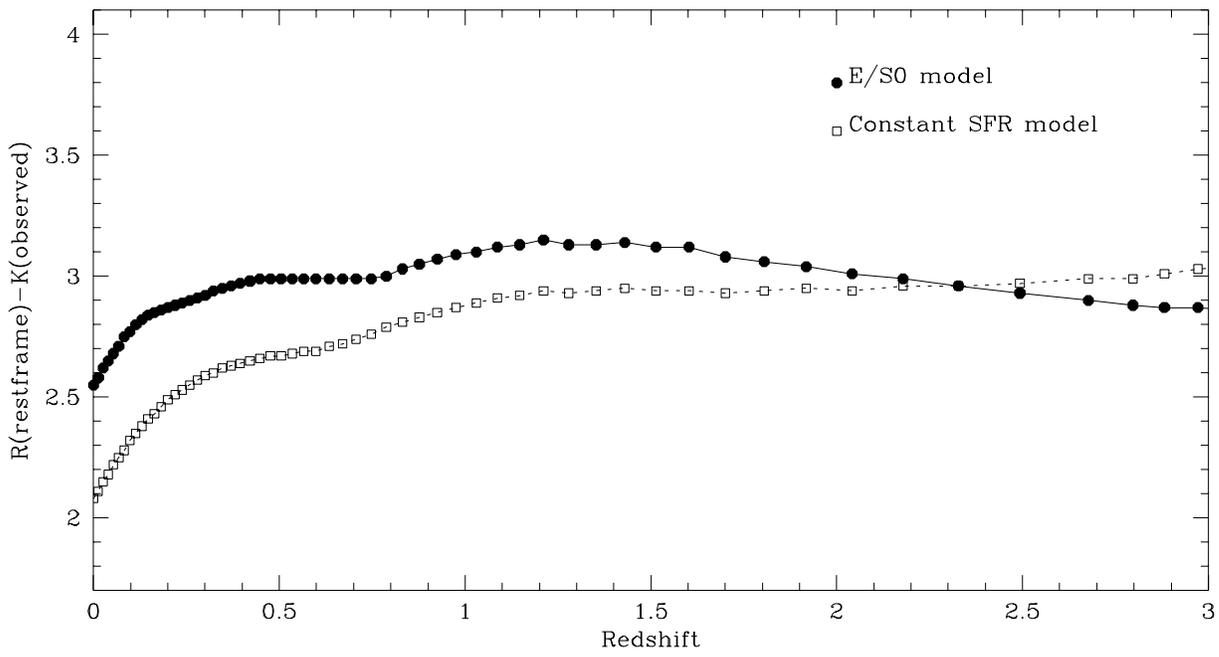
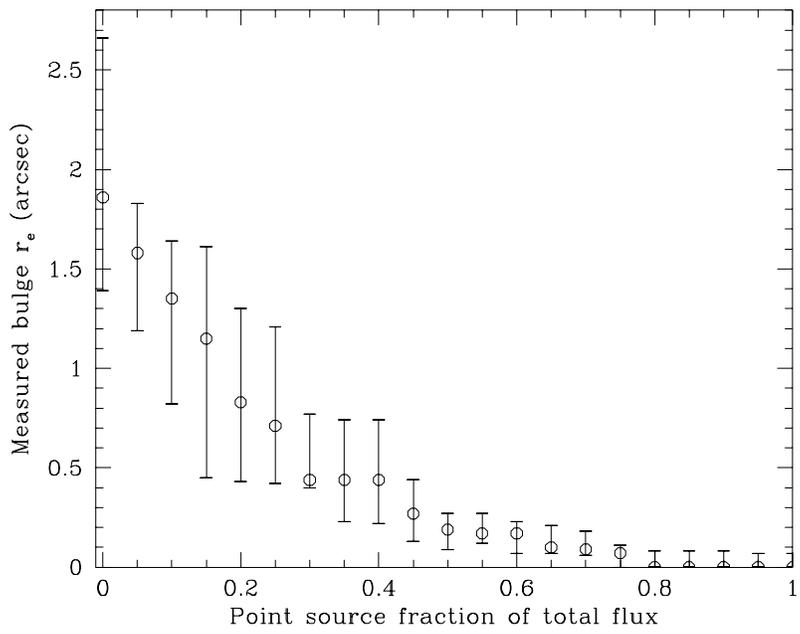
\end